\documentclass[a4paper,aps,twocolumn,nofootinbib]{revtex4}
\RequirePackage[english]{babel}
\RequirePackage[latin1]{inputenc}
\RequirePackage[T1]{fontenc}
\RequirePackage{mathrsfs}
\RequirePackage{amsmath}
\RequirePackage{amssymb}
\RequirePackage{amsbsy}
\RequirePackage{color}
\RequirePackage{bm}
\begin{document}
%%%%%%%%%%%%%%%%%%%%%%%%%%%%%%%%%%%%%%%%%%%%%%%%%%%%%%%%%%%%%%%%%%%%%%%%%%%%%%%%%%%%%%%%%%%%%%%%%%%
\title{\bf{General Dynamics of Spinors}}
\author{Luca Fabbri}
\affiliation{DIME, Universit\`{a} di Genova, P.Kennedy Pad.D, 16129 Genova, ITALY}
\date{\today}
%%%%%%%%%%%%%%%%%%%%%%%%%%%%%%%%%%%%%%%%%%%%%%%%%%%%%%%%%%%%%%%%%%%%%%%%%%%%%%%%%%%%%%%%%%%%%%%%%%%
\begin{abstract}
In this paper, we consider a general twisted-curved space-time hosting Dirac spinors and we take into account the Lorentz covariant polar decomposition of the Dirac spinor field: the corresponding decomposition of the Dirac spinor field equation leads to a set of field equations that are real and where spinorial components have disappeared while still maintaining Lorentz covariance. We will see that the Dirac spinor will contain two real scalar degrees of freedom, the module and the so-called Yvon-Takabayashi angle, and we will display their field equations. This will permit us to study the coupling of curvature and torsion respectively to the module and the YT angle.
\end{abstract}
%%%%%%%%%%%%%%%%%%%%%%%%%%%%%%%%%%%%%%%%%%%%%%%%%%%%%%%%%%%%%%%%%%%%%%%%%%%%%%%%%%%%%%%%%%%%%%%%%%%
\maketitle
%%%%%%%%%%%%%%%%%%%%%%%%%%%%%%%%%%%%%%%%%%%%%%%%%%%%%%%%%%%%%%%%%%%%%%%%%%%%%%%%%%%%%%%%%%%%%%%%%%%
\section{History}
The Dirac equation is one of the most impressive successes in all of physics (and as far as we can tell, in all of human achievements): conceived from the purely theoretical (or in Dirac's thoughts, aesthetic) reason to be a covariant first-order derivative field equation, it turned out to account for spin and matter/antimatter duality.

Such an extensively comprehensive description comes at the cost of a rather complicated formalism: as a start, spinors (in this paper we only consider Dirac spinors) are $4$-dimensional columns of complex scalar fields, amounting to $8$ real components. Moreover, the spinor formalism does not put in evidence the essence of any of the various components of a spinor field --- So is there a way in which to write the spinor formalism so that all components display a clear meaning? Also, can we reduce the variety of the components by proving that some of them are not in fact true degrees of freedom? And if yes, then how many degrees of freedom are actually present in a spinor?

These are all legitimate questions that researchers have been trying to answer, though not with the same impetus with which research has been done in more fashionable branches; still, some research has been done, and to our knowledge, the first to work on this problem were Jakobi and Lochak \cite{j-l}, followed after some time, but with a much richer research production, by Hestenes \cite{h1,h2,h3}.

The idea they had was to write the Dirac spinor field in the polar form: as a complex scalar can be written as the product of module times unitary phase, similarly the complex spinor should be writable as a column with four components, each of which being the product of a module times a unitary phase; while for scalars this construction is always trivial, spinors are defined in such a way that a spinorial transformation mixes the various components, and care must be exercised if we want the polar form to respect Lorentz symmetries. The works mentioned above do precisely this: they expound the spinor in a form that is polar while displaying a manifest Lorentz symmetry in its structure. As we will discuss later on, the polar form allows us to give a clear interpretation of the components of the spinor field and it shows which ones are artifacts and which ones are real degrees of freedom.

On the basis of these results, Hestenes went further to discuss \textit{zitterbewegung} effects in quantum theories \cite{h4,h5}.

As the polar form of scalars is used in the Schr\"{o}dinger equation to give the so-called Madelung decomposition, similarly we may take the polar form of spinors into the Dirac equation to perform an analogue of the Madelung decomposition: as above, such a decomposition is trivial for scalars, where only the splitting in real and imaginary parts is demanded, but for spinors we have to account for the fact that, beyond splitting real and imaginary parts, one also has to split the various components, and this in general entails a loss of manifest covariance; to overcome this, and maintain manifest covariance, one may employ a Gordon decomposition of the Dirac equation instead of the Dirac equation itself. Therefore, the final recipe goes as it follows: first, plug the polar form of the spinor into the Dirac spinor equation; then, multiply on the left by Clifford matrices and by the adjoint spinor to get scalar equations; finally, split real and imaginary parts to obtain real scalar equations. These equations will be manifestly covariant and real, but in polar decomposition \cite{k}.

In \cite{k} one may also find a study of the solutions based on a perturbative expansion in the Compton length.

Repeating the above recipe for all the linearly independent Clifford matrices would allow one to get all the independent decompositions. As we will see in what follows, some of them are already known, and remarkable: a first tells us that the Dirac Lagrangian on shell is identically zero; a second is the continuity equation for the velocity density, a third the partially-conserved axial-vector current for the spin density... the list goes on. Clearly, if this procedure is done exhaustively, one should expect that all these covariant real polar decompositions be altogether equivalent to the original Dirac spinor field equation.

This will be the case, as we are going to discuss later.

An interesting fact is that, in this polar decomposition, spinors can be written in a form that puts in evidence relevant information: the spinor thusly decomposed is given in terms of the $3$ components of its spin and the $3$ components of its velocity, plus its module and an additional quantity known in literature as the Yvon-Takabayashi angle, amounting to the expected $8$ components. Obviously, not all $8$ components are degrees of freedom, because one may perform boosts and rotations transferring the information about spin and velocity into the frame. Only the module and the Yvon-Takabayashi angle are degrees of freedom, encoding information about the dynamics, but while the interpretation of the module as what describes the density is well established, no interpretation for the YT angle is known so far, and not for lack of trying.

The first who tried some interpretation of the YT angle was de Broglie, whose failure in understanding it led him to dub the angle as ``mysterious''. But de Broglie's failures were very well justified, since the YT angle has remained no less mysterious ever since: in particular, Hestenes himself, in his papers on the polar decomposition, constantly raises the same question about this ``curious'' angle.

The most we could find was that in the very first of all these works \cite{h1}, Hestenes tries an interpretation of such a weird YT angle as something that might be connected to re-normalization: he writes that ``insofar as the problem of re-normalization is to calculate the re-distribution of charge due to interaction, it is the problem of calculating the YT angle'' (page 806, left column); on the other hand, in none of the subsequent papers is such an interpretation reconsidered, nor deepened, so far as we are aware.

Some direct reading of the polar decomposition may be done to argue that the YT angle could be related to what measures the mixture of particle/antiparticle degrees of freedom within a single spinor. On the other hand, we do not think this could be a viable interpretation, at least in the original form, for the reason that follows.

As already mentioned, when the polar decomposition is plugged into the field equations we obtain a corresponding polar decomposition of the field equations, which can then be further decomposed according to the usual Gordon procedure and then be split into real and imaginary parts: after this is done, it is possible to see that the YT angle is governed by its own field equation as it should be expected since it is a real degree of freedom. However, this means that the YT is a genuine dynamical quantity.

As such, it is expected to vary even in the free case, so that electrons would convert into positrons even without interactions. This does not seem to be what happens.

Maybe the truth is in between: the YT angle describes some dynamical process and not a mixture between particles and antiparticles, but \emph{symmetries} of the YT angle map particles into antiparticles. We do not know.

And in any case, the nature of the dynamical processes involving the YT angle would still escape us.

Yet another interesting consequence of the decomposed field equations, which are first-order derivative, is that a few of them can be combined, giving second-order derivative field equations of the Hamilton-Jacobi type as it has been described by Rodrigues and co-workers \cite{r-w,m-r-w}.

In this form, it is possible to assign other meanings to the YT angle, such as a corrective term to the mass of particles. In this sense, Rodrigues and co-workers might be hooking somehow to the initial idea of Hestenes about re-normalization, specifically mass re-normalization.

Another intriguing study to pursue would be the effect of torsion on the YT angle: the YT angle is intrinsically connected to spin, which is the source of torsion, and as a consequence the presence of torsion must have an impact on the dynamics of the Yvon-Takabayashi angle itself.

Some genuinely geometrical approach of the problem in space-times that are curved and twisted has been made by Rodrigues and co-workers, like for instance in \cite{Rodrigues:1996tj}.

However, we feel that the problem of the polar decomposition of spinor fields and spinor field equations, their full coupling, and the ensuing dynamics, have not yet received a systematic treatment. References are few and a little sparse over decades, and while the work of Hestenes is thorough, but it never considers the full coupling to the curved-twisted space-time, the work of Rodrigues and co-workers does, but consequences are not discussed much.

Instead we think that the relationships between torsion and the Yvon-Takabayashi angle should be investigated more deeply, and this is what we will do in this paper.
%%%%%%%%%%%%%%%%%%%%%%%%%%%%%%%%%%%%%%%%%%%%%%%%%%%%%%%%%%%%%%%%%%%%%%%%%%%%%%%%%%%%%%%%%%%%%%%%%%%
%%%%%%%%%%%%%%%%%%%%%%%%%%%%%%%%%%%%%%%%%%%%%%%%%%%%%%%%%%%%%%%%%%%%%%%%%%%%%%%%%%%%%%%%%%%%%%%%%%%
\section{Fundamental Setting}
We will begin by recalling the results presented in the introductory part, but we will follow the form used in the references \cite{Fabbri:2016msm} and \cite{Fabbri:2016laz}, and more in particular \cite{Fabbri:2016fxt}.

There are two reasons for this: the first is an accident, that is, as the results of Jakobi and Lochak were unknown to Hestenes, similarly the results of Jakobi and Lochak as well as of Hestenes were unknown to me at the moment of writing references \cite{Fabbri:2016msm,Fabbri:2016laz}, \cite{Fabbri:2016fxt} (while Jakobi and Lochak wrote their paper in French, thus giving a justification to Hestenes for not being aware of it, there is no excuse for my ignorance of any previous result); the other reason is more technical (and more noble), and it is that whereas Hestenes studies the polar decomposition by introducing a new formalism called space-time algebra, for the above references \cite{Fabbri:2016msm,Fabbri:2016laz,Fabbri:2016fxt} no new formalism is necessary.

Despite Hestenes way and our way of making the polar decomposition of spinor fields and spinor field equations put the decomposed spinor field equations in real form, nevertheless Hestenes way converts the use of the Clifford algebra into the use of another form of algebra, while our way leaves the spinor field equations in a form in which only tensors are present; in the perspective of rendering the polar decomposition an instrument to ease the visual recognition of the quantities involved, we believe that the form we use is the simplest that was ever employed.

So, to really start the summary of the previous results, we first remind the reader that for a general description of space-time, the metric is given by $g_{\alpha\rho}$ and it will be used to move coordinate indices; tetrads $e^{\alpha}_{a}$ are always taken to be ortho-normal $g_{\alpha\rho}e^{\alpha}_{a}e^{\rho}_{b}\!=\!\eta_{ab}$ and used to pass from coordinate (Greek) indices to Lorentz (Latin) indices: thus the Minkowskian matrix $\eta_{ab}$ is used to move the Lorentz indices. Matrices $\boldsymbol{\gamma}^{a}$ belong to the Clifford algebra, then we define $\left[\boldsymbol{\gamma}^{a}\!,\!\boldsymbol{\gamma}^{b}\right]\!=\! 4\boldsymbol{\sigma}^{ab}$ and $2i\boldsymbol{\sigma}_{ab}\!=\!\varepsilon_{abcd}\boldsymbol{\pi}\boldsymbol{\sigma}^{cd}$ in which the parity-odd matrix $\boldsymbol{\pi}$ is implicitly defined (this is what is usually indicated as gamma with index five, but since in the space-time this index has no meaning we prefer to use a notation with no index), and with $\boldsymbol{\gamma}_{0}$ we can define for spinor $\psi$ the conjugate spinor $\overline{\psi}\!=\!\psi^{\dagger}\boldsymbol{\gamma}_{0}$ such that
\begin{eqnarray}
&2i\overline{\psi}\boldsymbol{\sigma}^{ab}\psi\!=\!M^{ab}\\
&\overline{\psi}\boldsymbol{\gamma}^{a}\boldsymbol{\pi}\psi\!=\!S^{a}\\
&\overline{\psi}\boldsymbol{\gamma}^{a}\psi\!=\!U^{a}\\
&i\overline{\psi}\boldsymbol{\pi}\psi\!=\!\Theta\\
&\overline{\psi}\psi\!=\!\Phi
\end{eqnarray}
are all real: Clifford matrices verify
\begin{eqnarray}
&\boldsymbol{\gamma}_{i}\boldsymbol{\gamma}_{j}\boldsymbol{\gamma}_{k}
=\boldsymbol{\gamma}_{i}\eta_{jk}-\boldsymbol{\gamma}_{j}\eta_{ik}+\boldsymbol{\gamma}_{k}\eta_{ij}
+i\varepsilon_{ijkq}\boldsymbol{\pi}\boldsymbol{\gamma}^{q}
\end{eqnarray}
in general, then we can demonstrate that
\begin{eqnarray}
&M_{ab}\!=\!(\Phi^{2}\!+\!\Theta^{2})^{-1}(U^{j}S^{k}\varepsilon_{jkab}\Phi\!+\!U_{[a}S_{b]}\Theta)
\end{eqnarray}
showing that only the vector and axial-vector with scalar and pseudo-scalar are independent, and so by defining 
\begin{eqnarray}
&S^{a}\!=\!(\Phi^{2}\!+\!\Theta^{2})^{\frac{1}{2}}s^{a}\label{aux1}\\
&U^{a}\!=\!(\Phi^{2}\!+\!\Theta^{2})^{\frac{1}{2}}u^{a}\label{aux2}
\end{eqnarray}
as well as 
\begin{eqnarray}
&\Theta\!=\!2\phi^{2}\sin{\beta}\label{b2}\\
&\Phi\!=\!2\phi^{2}\cos{\beta}\label{b1}
\end{eqnarray}
we remain with the constraints
\begin{eqnarray}
&u_{a}u^{a}\!=\!-s_{a}s^{a}\!=\!1\label{norm}\\
&u_{a}s^{a}\!=\!0\label{orthogonal}
\end{eqnarray}
and only two scalar fields $\phi$ and $\beta$ as the true independent degrees of freedom. We also have the relationships
\begin{eqnarray}
&\!\!\!\!\psi\overline{\psi}\!\equiv\!\frac{1}{2}
\phi^{2}[(u_{a}\boldsymbol{\mathbb{I}}\!+\!s_{a}\boldsymbol{\pi})\boldsymbol{\gamma}^{a}
\!\!+\!e^{-i\beta\boldsymbol{\pi}}(\boldsymbol{\mathbb{I}}
\!-\!2u_{a}s_{b}\boldsymbol{\sigma}^{ab}\boldsymbol{\pi})]
\label{F}
\end{eqnarray}
which are valid in the most general of the circumstances.

From the metric we define $\Lambda^{\sigma}_{\alpha\nu}$ as the symmetric connection; with it $\Omega^{a}_{b\pi}\!=\!\xi^{\nu}_{b}\xi^{a}_{\sigma}(\Lambda^{\sigma}_{\nu\pi}\!-\!\xi^{\sigma}_{i}\partial_{\pi}\xi_{\nu}^{i})$ will be the spin connection in general. Hence it is possible to define
\begin{eqnarray}
&\boldsymbol{\Omega}_{\mu}
=\frac{1}{2}\Omega^{ab}_{\phantom{ab}\mu}\boldsymbol{\sigma}_{ab}
\!+\!iqA_{\mu}\boldsymbol{\mathbb{I}}\label{spinorialconnection}
\end{eqnarray}
in terms of the spin connection and the gauge potential of charge $q$ and called spinorial connection. Remark that because torsion is a tensor unrelated to any type of gauge transformation, then it need not be inside any connection and it can be kept separated away from it, and this is the reason why we can give the formalism in the torsionless case, adding torsion only in the dynamics. Furthermore, as discussed in the literature above and in the references therein, torsion can be considered to be completely antisymmetric, which means that in the space-time it is the dual of an axial-vector $W_{\sigma}$ as usually indicated.

Now, with this basic notation we may study the general form of spinor fields, and in papers \cite{j-l,h1,Fabbri:2016msm} it has been demonstrated that (in the case where at least one of the scalars $\Theta$ and $\Phi$ is non-zero) it is possible to find a frame where the most general spinor reduces to
\begin{eqnarray}
&\!\psi'\!=\!\phi\ e^{-\frac{i}{2}\beta\boldsymbol{\pi}}\!\left(\!\begin{tabular}{c}
$1$\\
$0$\\
$1$\\
$0$
\end{tabular}\!\right)
\label{spinorspecial}
\end{eqnarray}
up to a third-axis reflection and up to the transformation of discrete type $\psi\!\rightarrow\!\boldsymbol{\pi}\psi$ in general, and which is given in terms of the $\phi$ and $\beta$ degrees of freedom above (the case in which both $\Theta\!=\!\Phi\!=\!0$ is not treated in \cite{j-l}, it is quickly mentioned in \cite{h1} and it is thoroughly studied in \cite{Fabbri:2016msm}, but despite this case is very interesting and in fact quite well studied in the literature \cite{Vignolo:2011qt, daRocha:2013qhu, daSilva:2012wp, Ablamowicz:2014rpa, daRocha:2016bil, daRocha:2008we, Villalobos:2015xca, Cavalcanti:2014uta}, nevertheless there does not yet seem to be a general consensus about what these so-called singular spinors might describe, so we will leave their treatment aside in this article); a general spinorial transformation $\boldsymbol{S}^{-1}$ can be employed to go from this form back to the most general form of the spinor given by
\begin{eqnarray}
&\!\psi\!=\!\phi\sqrt{\frac{2}{\gamma+1}}e^{i\alpha}\left(\!\begin{tabular}{c}
$e^{\frac{i}{2}\beta}\left(\frac{\gamma+1}{2}\boldsymbol{\mathbb{I}}\!-\!
\gamma\vec{v}\!\cdot\!\vec{\frac{\boldsymbol{\sigma}}{2}}\right)\xi$\\
$e^{-\frac{i}{2}\beta}\left(\frac{\gamma+1}{2}\boldsymbol{\mathbb{I}}\!+\!
\gamma\vec{v}\!\cdot\!\vec{\frac{\boldsymbol{\sigma}}{2}}\right)\xi$
\end{tabular}\!\right)
\label{spinor}
\end{eqnarray}
up to the transformation of discrete type $\psi\!\rightarrow\!\boldsymbol{\pi}\psi$ and in which $\gamma\!=\!1/\!\sqrt{1\!-\!v^{2}}$ is the relativistic factor (which is not to be confused with the Clifford matrices) given in terms of the velocity $\vec{v}$ while $\xi$ such that $\xi^{\dagger}\xi\!=\!1$ is an arbitrary semi-spinor and $\alpha$ is a generic unitary phase: spinors are consequently writable in the most general circumstance in this form, which has the advantage of displaying what are the degrees of freedom and what components can be removed with suitable local spinor transformations, thus reducing the spinor to the form (\ref{spinorspecial}) above. With a direct calculation one can demonstrate that the directions are
\begin{eqnarray}
&s^{a}\!=\!\frac{1}{\gamma+1}\left(\!\begin{tabular}{c}
$2\gamma(\gamma\!+\!1)(\vec{v}\!\cdot\!\vec{\varsigma})$\\
$2(\gamma+1)\vec{\varsigma}\!+\!2\gamma^{2}(\vec{v}\!\cdot\!\vec{\varsigma})\vec{v}$
\end{tabular}\!\right)\label{s}\\
&u^{a}\!=\!\frac{1}{\gamma+1}\left(\!\begin{tabular}{c}
$\frac{1}{2}(\gamma\!+\!1)^{2}+\frac{1}{2}\gamma^{2}(\vec{v}\!\cdot\!\vec{v})$\\
$\gamma(\gamma\!+\!1)\vec{v}$
\end{tabular}\!\right)\label{u}
\end{eqnarray}
where $\xi^{\dagger}\vec{\boldsymbol{\sigma}}\xi\!=\!2\vec{\varsigma}$ is the spin and with scalar and pseudo-scalar being given by (\ref{b2}, \ref{b1}): this is expected since after all $\phi$ and $\beta$ are scalars. This is the form of the spinor field that we are going to employ in the rest of the work.

Computing the derivative of (\ref{spinorspecial}) and applying $\boldsymbol{S}^{-1}$ or computing the spinorial covariant derivative of (\ref{spinor}) gives
\begin{eqnarray}
\nonumber
&\boldsymbol{\nabla}_{\mu}\psi\!=\![\nabla_{\mu}\ln{\phi}\mathbb{I}
\!-\!\frac{i}{2}\nabla_{\mu}\beta\boldsymbol{\pi}+\\
&+i(qA_{\mu}\!-\!P_{\mu})\mathbb{I}
\!+\!\frac{1}{2}(\Omega_{ij\mu}\!-\!R_{ij\mu})\boldsymbol{\sigma}^{ij}]\psi
\label{decspinder}
\end{eqnarray}
in which $\boldsymbol{S}^{-1}\partial_{\mu}\boldsymbol{S}\!=\!iP_{\mu}\mathbb{I}
+\frac{1}{2}R_{ij\mu}\boldsymbol{\sigma}^{ij}$ with $P_{\mu}$ gauge vector and $R_{\alpha\nu\mu}$ spin connection, $A_{\mu}$ and $\Omega_{ij\mu}$ being the usual electrodynamic potential and gravitational strength: the fact that the difference of two connections is a tensor and that $\phi$ and $\beta$ are scalars ensures that the above spinorial covariant derivative is indeed covariant, and in addition it is such in each term separately. From this form we get
\begin{eqnarray}
&\nabla_{\mu}s_{\alpha}\!=\!(R\!-\!\Omega)_{\rho \alpha\mu}s^{\rho}\label{ds}\\
&\nabla_{\mu}u_{\alpha}\!=\!(R\!-\!\Omega)_{\rho\alpha\mu}u^{\rho}\label{du}
\end{eqnarray}
from which we can also calculate all the divergences and curls of these vectors in every equation that follows.

Expression (\ref{spinor}) is the polar decomposition of the Dirac spinor field: in general, the spinor field possesses a total number of $8$ real components, and they are given by the spin axial-vector $s^{a}$ and velocity vector $u^{a}$ with the module given by the scalar $\phi$ and the Yvon-Takabayashi angle given by pseudo-scalar $\beta$ as in (\ref{b2}, \ref{b1}); despite amounting to a total of $10$ real components, not all of them are independent, as it can be appreciated from the existence of constraints (\ref{norm}, \ref{orthogonal}), or the fact that according to the expressions (\ref{s}, \ref{u}) only $\vec{\varsigma}$ and $\vec{v}$ are needed. Within the spinor (\ref{spinor}) the $3$ real components of $\vec{\varsigma}$ are codified inside the $2$-dimensional column of complex-valued $\xi$ subject to the $\xi^{\dagger}\xi\!=\!1$ constraint; the phase $\alpha$ does not constitute a real degree of freedom since by performing some Lorentz transformation it is always possible to remove it, whether the spinor is charged or neutral \cite{Fabbri:2016msm}. The only two scalar real degrees of freedom are the module and the YT angle.

Expression (\ref{decspinder}) shows that whenever some local spinorial transformation transfers components from the spinor to the frame, the derivatives of those components, lost as derivatives of the spinor, are gained as components of the connection: thus, the whole spinorial covariant derivative is covariant even for local transformations. Indeed this is what we should have expected, and we shall employ this spinorial covariant derivative to study the dynamics. 

For the dynamics we will employ the action, or system of field equations, we have given in \cite{Fabbri:2016laz} and \cite{Fabbri:2016fxt}.

As compared to the works of Hestenes \cite{h2,h3} and also to that of Krueger \cite{k}, the dynamical system of field equations presented in \cite{Fabbri:2016laz,Fabbri:2016fxt} is more complete, as it accounts beside electrodynamics also for gravity with torsion.

For the Dirac spinor field equations, we consider those with the most general coupling to the axial-vector torsion which are given according to the following form
\begin{eqnarray}
&i\boldsymbol{\gamma}^{\mu}\boldsymbol{\nabla}_{\mu}\psi
\!-\!XW_{\sigma}\boldsymbol{\gamma}^{\sigma}\boldsymbol{\pi}\psi\!-\!m\psi\!=\!0
\label{dfe}
\end{eqnarray}
where the $X$ is the torsion-spin coupling constant and in which $m$ is of course the mass of the spinor field.

Having the polar decomposition of the spinor field and the spinor field equations, we may plug (\ref{spinor}) or directly its derivative (\ref{decspinder}) into (\ref{dfe}) getting the polar decomposition of the spinor field equations, and then we might proceed in splitting real and imaginary parts as in the Madelung decomposition; but as we have already mentioned, such a procedure for spinors violates manifest covariance of the equations unless additionally also the Gordon decomposition is performed: the order in which these three steps must be performed is to start with the Gordon decompositions, that is having the spinor field equations multiplied by all matrices $\mathbb{I}, \boldsymbol{\pi}, \boldsymbol{\gamma}^{a}, \boldsymbol{\gamma}^{a}\boldsymbol{\pi}, \boldsymbol{\sigma}^{ab}$ and the conjugate spinor field, and then splitting imaginary and real parts, so that the resulting $10$ real tensorial equations are given respectively by the following list of field equations
\begin{eqnarray}
&\frac{i}{2}(\overline{\psi}\boldsymbol{\gamma}^{\mu}\boldsymbol{\nabla}_{\mu}\psi
\!-\!\boldsymbol{\nabla}_{\mu}\overline{\psi}\boldsymbol{\gamma}^{\mu}\psi)
\!-\!XW_{\sigma}S^{\sigma}\!-\!m\Phi\!=\!0\\
&\nabla_{\mu}U^{\mu}\!=\!0
\end{eqnarray}
\begin{eqnarray}
&\frac{i}{2}(\overline{\psi}\boldsymbol{\gamma}^{\mu}\boldsymbol{\pi}\boldsymbol{\nabla}_{\mu}\psi
\!-\!\boldsymbol{\nabla}_{\mu}\overline{\psi}\boldsymbol{\gamma}^{\mu}\boldsymbol{\pi}\psi)
\!-\!XW_{\sigma}U^{\sigma}\!=\!0\\
&\nabla_{\mu}S^{\mu}\!-\!2m\Theta\!=\!0
\end{eqnarray}
\begin{eqnarray}
\nonumber
&\frac{i}{2}(\overline{\psi}\boldsymbol{\nabla}^{\alpha}\psi
\!-\!\boldsymbol{\nabla}^{\alpha}\overline{\psi}\psi)
\!-\!\frac{1}{2}\nabla_{\mu}M^{\mu\alpha}-\\
&-\frac{1}{2}XW_{\sigma}M_{\mu\nu}\varepsilon^{\mu\nu\sigma\alpha}\!-\!mU^{\alpha}\!=\!0
\label{vr}\\
\nonumber
&\nabla_{\alpha}\Phi
\!-\!2(\overline{\psi}\boldsymbol{\sigma}_{\mu\alpha}\!\boldsymbol{\nabla}^{\mu}\psi
\!-\!\!\boldsymbol{\nabla}^{\mu}\overline{\psi}\boldsymbol{\sigma}_{\mu\alpha}\psi)+\\
&+2X\Theta W_{\alpha}\!=\!0\label{vi}
\end{eqnarray}
\begin{eqnarray}
\nonumber
&\nabla_{\nu}\Theta\!-\!
2i(\overline{\psi}\boldsymbol{\sigma}_{\mu\nu}\boldsymbol{\pi}\boldsymbol{\nabla}^{\mu}\psi\!-\!
\boldsymbol{\nabla}^{\mu}\overline{\psi}\boldsymbol{\sigma}_{\mu\nu}\boldsymbol{\pi}\psi)-\\
&-2X\Phi W_{\nu}\!+\!2mS_{\nu}\!=\!0\label{ar}\\
\nonumber
&(\boldsymbol{\nabla}_{\alpha}\overline{\psi}\boldsymbol{\pi}\psi
\!-\!\overline{\psi}\boldsymbol{\pi}\boldsymbol{\nabla}_{\alpha}\psi)
\!-\!\frac{1}{2}\nabla^{\mu}M^{\rho\sigma}\varepsilon_{\rho\sigma\mu\alpha}+\\
&+2XW^{\mu}M_{\mu\alpha}\!=\!0\label{ai}
\end{eqnarray}
\begin{eqnarray}
\nonumber
&\nabla^{\mu}S^{\rho}\varepsilon_{\mu\rho\alpha\nu}
\!+\!i(\overline{\psi}\boldsymbol{\gamma}_{[\alpha}\!\boldsymbol{\nabla}_{\nu]}\psi
\!-\!\!\boldsymbol{\nabla}_{[\nu}\overline{\psi}\boldsymbol{\gamma}_{\alpha]}\psi)+\\
&+2XW_{[\alpha}S_{\nu]}\!=\!0\\
\nonumber
&\nabla^{[\alpha}U^{\nu]}\!+\!i\varepsilon^{\alpha\nu\mu\rho}
(\overline{\psi}\boldsymbol{\gamma}_{\rho}\boldsymbol{\pi}\!\boldsymbol{\nabla}_{\mu}\psi\!-\!\!
\boldsymbol{\nabla}_{\mu}\overline{\psi}\boldsymbol{\gamma}_{\rho}\boldsymbol{\pi}\psi)-\\
&-2XW_{\sigma}U_{\rho}\varepsilon^{\alpha\nu\sigma\rho}\!-\!2mM^{\alpha\nu}\!=\!0
\end{eqnarray}
where the polar decomposition must be done. Of course, one possibility would be to proceed by brute force, that is polar decomposing all of them, but in doing so we would simply end up having an enormous amount of terms from which little insight can be gained; instead, we will make the polar decomposition only for the four vectorial type of equations (\ref{vr}, \ref{vi}, \ref{ar}, \ref{ai}), getting the expressions
\begin{eqnarray}
\nonumber
&-\nabla_{\mu}\ln{\phi}M^{\mu\sigma}
\!+\!\frac{1}{2}(\frac{1}{2}\nabla_{\mu}\beta\!-\!XW_{\mu})M_{\pi\nu}
\varepsilon^{\pi\nu\mu\sigma}-\\
\nonumber
&-(qA\!-\!P)^{\sigma}\Phi\!-\!\frac{1}{8}(\Omega\!-\!R)^{\alpha\nu\rho}M_{\pi\kappa}
\varepsilon_{\alpha\nu\rho\mu}\varepsilon^{\pi\kappa\sigma\mu}+\\
&+\frac{1}{2}(\Omega\!-\!R)_{\mu a}^{\phantom{\mu a}a}M^{\mu\sigma}\!-\!mU^{\sigma}\!=\!0\\
\nonumber
&-\nabla_{\sigma}\ln{\phi}\Phi\!+\!(\frac{1}{2}\nabla_{\sigma}\beta\!-\!XW_{\sigma})\Theta+\\
\nonumber
&+(qA\!-\!P)^{\mu}M_{\mu\sigma}
\!+\!\frac{1}{4}(\Omega\!-\!R)^{\alpha\nu\rho}\varepsilon_{\alpha\nu\rho\sigma}\Theta+\\
&+\frac{1}{2}(\Omega\!-\!R)_{\sigma a}^{\phantom{\sigma a}a}\Phi\!=\!0\label{int1}
\end{eqnarray}
\begin{eqnarray}
\nonumber
&\nabla_{\sigma}\ln{\phi}\Theta\!+\!(\frac{1}{2}\nabla_{\sigma}\beta\!-\!XW_{\sigma})\Phi-\\
\nonumber
&-\frac{1}{2}(qA\!-\!P)^{\mu}M^{\pi\kappa}\varepsilon_{\pi\kappa\mu\sigma}
\!+\!\frac{1}{4}(\Omega\!-\!R)^{\alpha\nu\rho}\varepsilon_{\alpha\nu\rho\sigma}\Phi-\\
&-\frac{1}{2}(\Omega\!-\!R)_{\sigma a}^{\phantom{\sigma a}a}\Theta\!+\!mS_{\sigma}
\!=\!0\label{int2}\\
\nonumber
&\frac{1}{2}\nabla_{\mu}\ln{\phi}M_{\pi\kappa}\varepsilon^{\pi\kappa\mu\sigma}
\!+\!(\frac{1}{2}\nabla_{\mu}\beta\!-\!XW_{\mu})M^{\mu\sigma}+\\
\nonumber
&+(qA\!-\!P)^{\sigma}\Theta
\!+\!\frac{1}{4}(\Omega\!-\!R)^{\alpha\nu\rho}M^{\mu\sigma}\varepsilon_{\alpha\nu\rho\mu}-\\
&-\frac{1}{4}(\Omega\!-\!R)_{\mu a}^{\phantom{\mu a}a}M_{\pi\kappa}
\varepsilon^{\pi\kappa\mu\sigma}\!=\!0
\end{eqnarray}
where further simplifications can occur. For instance, we may substitute the bi-linear antisymmetric tensor, then have the bi-linear vector and axial-vector normalized in equations (\ref{int1}, \ref{int2}), thus obtaining the final equations
\begin{eqnarray}
\nonumber
&\frac{1}{2}\nabla_{\alpha}\ln{\phi^{2}}\cos{\beta}
\!-\!(\frac{1}{2}\nabla_{\alpha}\beta\!-\!XW_{\alpha})\sin{\beta}+\\
\nonumber
&+(P\!-\!qA)^{\mu}(u^{\rho}s^{\sigma}\varepsilon_{\rho\sigma\mu\alpha}\cos{\beta}
\!+\!u_{[\mu}s_{\alpha]}\sin{\beta})-\\
\nonumber
&-\frac{1}{2}(\Omega\!-\!R)_{\alpha\mu}^{\phantom{\alpha\mu}\mu}\cos{\beta}-\\
&-\frac{1}{4}(\Omega\!-\!R)^{\rho\sigma\mu}\varepsilon_{\rho\sigma\mu\alpha}\sin{\beta}\!=\!0
\label{a1}\\
\nonumber
&\frac{1}{2}\nabla_{\nu}\ln{\phi^{2}}\sin{\beta}
\!+\!(\frac{1}{2}\nabla_{\nu}\beta\!-\!XW_{\nu})\cos{\beta}+\\
\nonumber
&+(P\!-\!qA)^{\mu}(u^{\rho}s^{\sigma}\varepsilon_{\rho\sigma\mu\nu}\sin{\beta}\!-\!u_{[\mu}s_{\nu]}\cos{\beta})+\\
\nonumber
&+\frac{1}{4}(\Omega\!-\!R)^{\rho\sigma\mu}\varepsilon_{\rho\sigma\mu\nu}\cos{\beta}-\\
&-\frac{1}{2}(\Omega\!-\!R)_{\nu\mu}^{\phantom{\nu\mu}\mu}\sin{\beta}\!+\!ms_{\nu}\!=\!0
\label{a2}
\end{eqnarray}
which still seem quite far from being easy to manipulate.

However, after they are diagonalized, they result into
\begin{eqnarray}
\nonumber
&\frac{1}{2}\varepsilon_{\mu\alpha\nu\iota}(R\!-\!\Omega)^{\alpha\nu\iota}
\!-\!2(P\!-\!qA)^{\iota}u_{[\iota}s_{\mu]}-\\
&-2XW_{\mu}\!+\!\nabla_{\mu}\beta\!+\!2s_{\mu}m\cos{\beta}\!=\!0\label{f1}\\
\nonumber
&(R\!-\!\Omega)_{\mu a}^{\phantom{\mu a}a}
\!-\!2(P\!-\!qA)^{\rho}u^{\nu}s^{\alpha}\varepsilon_{\mu\rho\nu\alpha}+\\
&+2s_{\mu}m\sin{\beta}\!+\!\nabla_{\mu}\ln{\phi^{2}}\!=\!0\label{f2}
\end{eqnarray}
which are manifestly covariant and clearly real, and they are the polar decomposition of the spinor field equations.

At this point the reader might feel justified in thinking that we have been lazy, since we have started from a total of ten Madelung-Gordon decompositions but we ended having only two, while on the contrary the truth is that in doing so we have been extremely efficient, because these are the only two equations needed: in fact, it is possible to prove that (\ref{f1}, \ref{f2}) do imply all the remaining equations, because in general (\ref{f1}, \ref{f2}) are equivalent to the original spinor field equations, as it was demonstrated in \cite{Fabbri:2016laz}.

Equations (\ref{f1}, \ref{f2}) are field equations for the two scalar fields $\phi$ and $\beta$ determining all their derivatives and therefore amounting to the $2\!\times\!4\!=\!8$ parts of the original spinor field equations: consequently, such a polar decomposition converts a spinor field equation into two vector equations for the two scalar real degrees of freedom. And of course the result is manifestly covariant and obviously real.

We believe this to be a nice result. Most of the papers by Hestenes and Krueger on the subject consist in seeking to perform all thinkable decompositions and projections in order to recover the full list of essential equations, but so far as we are aware they never pointed out what is the core of equations from which all other equations can be derived, and it is quite satisfying that in the end such a core is given by a pair of very simple equations indeed.

Field equations (\ref{f1}, \ref{f2}), however, consist of a mixture of various elements, while a cleaner form is obtained for second-order derivative field equations. To go to a higher order of derivation without burdening the formalism we define the pair of vectorial type of potentials
\begin{eqnarray}
&\!G_{\mu}\!=\!-(R\!-\!\Omega)_{\mu a}^{\phantom{\mu a}a}
\!+\!2(P\!-\!qA)^{\rho}u^{\nu}s^{\alpha}\varepsilon_{\mu\rho\nu\alpha}\\
&\!\!K_{\mu}\!=\!2XW_{\mu}
\!-\!\frac{1}{2}\varepsilon_{\mu\alpha\nu\iota}(R\!-\!\Omega)^{\alpha\nu\iota}
\!+\!2(P\!-\!qA)^{\iota}u_{[\iota}s_{\mu]}
\end{eqnarray}
given in terms of all basic fields: the first couple of these second-order derivative field equations is given by
\begin{eqnarray}
\nonumber
&\phi^{-2}\nabla^{2}\phi^{2}\!+\!(2m)^{2}+\\
&+2ms^{\mu}(G_{\mu}\sin{\beta}\!+\!K_{\mu}\cos{\beta})\!-\!(\nabla_{\mu}G^{\mu}\!+\!G^{2})
\!=\!0\label{KG}\\
\nonumber
&\nabla^{2}\beta\!-\!(2m)^{2}\sin{\beta}\cos{\beta}-\\
&-2ms^{\mu}(G_{\mu}\cos{\beta}\!+\!K_{\mu}\sin{\beta})\!-\!\nabla_{\mu}K^{\mu}
\!=\!0\label{sKG}
\end{eqnarray}
the first as a Klein-Gordon equation for the scalar of real mass and the second as a sine--Klein-Gordon equation for the pseudo-scalar of imaginary mass; still cleaner are
\begin{eqnarray}
&\!\nabla_{\mu}(\phi^{2}\nabla^{\mu}\frac{\beta}{2})
\!-\!\frac{1}{2}(\nabla_{\mu}K^{\mu}\!+\!K_{\mu}G^{\mu})\phi^{2}\!=\!0\label{cont}\\
&\!\!\!\!\left|\nabla \frac{\beta}{2}\right|^{2}\!\!-\!m^{2}\!-\!\phi^{-1}\nabla^{2}\phi
\!+\!\frac{1}{2}(\nabla_{\mu}G^{\mu}\!+\!\frac{1}{2}G^{2}\!-\!\frac{1}{2}K^{2})\!=\!0\label{HJ}
\end{eqnarray}
as a continuity equation and a Hamilton-Jacobi equation in which the field $\beta/2$ is a parity-odd action functional whereas $\phi^{-1}\nabla^{2}\phi$ is the quantum potential \cite{r-w,m-r-w}.

Finally the geometrical field equations are given by
\begin{eqnarray}
&\nabla_{\alpha}(\partial W)^{\alpha\mu}\!+\!M^{2}W^{\mu}\!=\!2X\phi^{2}s^{\mu}
\end{eqnarray}
with
\begin{eqnarray}
&\nabla_{\sigma}F^{\sigma\mu}\!=\!2q\phi^{2}u^{\mu}
\end{eqnarray}
and
\begin{eqnarray}
\nonumber
&R^{\rho\sigma}\!-\!\frac{1}{2}Rg^{\rho\sigma}\!-\!\Lambda g^{\rho\sigma}
\!=\!\frac{k}{2}[M^{2}(W^{\rho}W^{\sigma}\!\!-\!\!\frac{1}{2}W^{\alpha}W_{\alpha}g^{\rho\sigma})+\\
\nonumber
&+\frac{1}{4}(\partial W)^{2}g^{\rho\sigma}
\!-\!(\partial W)^{\sigma\alpha}(\partial W)^{\rho}_{\phantom{\rho}\alpha}+\\
\nonumber
&+\frac{1}{4}F^{2}g^{\rho\sigma}\!-\!F^{\rho\alpha}\!F^{\sigma}_{\phantom{\sigma}\alpha}-\\
\nonumber
&-\phi^{2}[(XW\!-\!\nabla\frac{\beta}{2})^{\sigma}s^{\rho}
\!+\!(XW\!-\!\nabla\frac{\beta}{2})^{\rho}s^{\sigma}]-\\
\nonumber
&-\phi^{2}[(qA\!-\!P)^{\sigma}u^{\rho}\!+\!(qA\!-\!P)^{\rho}u^{\sigma}]+\\
&+\frac{1}{4}\phi^{2}[(\Omega\!-\!R)_{ij}^{\phantom{ij}\sigma}\varepsilon^{\rho ijk}
\!+\!(\Omega\!-\!R)_{ij}^{\phantom{ij}\rho}\varepsilon^{\sigma ijk}]s_{k}]
\end{eqnarray}
and these field equations have the most general validity.

This concludes the introduction of the most important identities and field equations, which are at the same time terrific and terrifying: they are terrific because by having the spinor field equations re-written in a polar form where all degrees of freedom are real and covariant, we reduced the study of spinors to the study of real tensors; however, they are terrifying because despite having simplified so much, and indeed as much as it is thinkable, the resulting equations still seem far from being easy to manipulate.

So in what is next, we begin to set assumptions.
%%%%%%%%%%%%%%%%%%%%%%%%%%%%%%%%%%%%%%%%%%%%%%%%%%%%%%%%%%%%%%%%%%%%%%%%%%%%%%%%%%%%%%%%%%%%%%%%%%%
%%%%%%%%%%%%%%%%%%%%%%%%%%%%%%%%%%%%%%%%%%%%%%%%%%%%%%%%%%%%%%%%%%%%%%%%%%%%%%%%%%%%%%%%%%%%%%%%%%%
\section{Some Special Cases}
One of the most ubiquitous terms that has appeared is the term $\Omega-\!R$ measuring the presence of the gravitational field after having subtracted the information related to the spinor frame; as things seem to be, one might assume that this difference encodes the pure gravitational information, and it should vanish if no gravitation is present at all: as a matter of fact it has been proven in \cite{Fabbri:2016fxt} that in absence of gravity it is possible to arrange tetrads in such a way that they may cancel all information related to the spinor frame, even if the spinor has a precession.

Consequently, and for the sake of simplicity, we will be assuming that $\Omega-\!R\!=\!0$ identically in what follows.

In non-gravitational case, identities and field equations undergo to a spectacular reduction: as a start we have
\begin{eqnarray}
&\nabla_{\mu}s_{\alpha}\!=\!\nabla_{\mu}u_{\alpha}\!=\!0
\end{eqnarray}
which tells us that in absence of gravity the vector fields are covariantly constant; in particular, the velocity vector is constant, and furthermore it is divergenceless, meaning that the Dirac field, seen as a fluid, is incompressible.

Because in nature there exist fields that are neutral, it is possible to wonder what happens when we study these fields, so we will assume $q\!=\!0$ as well as $P\!=\!0$ in order to simplify all the calculations as much as it is possible.

And as another hypothesis, we will begin to assess the instance that is given by the totally free field.

The Dirac equations (\ref{f1}, \ref{f2}) in the free case are
\begin{eqnarray}
&\nabla_{\mu}\beta/2\!+\!s_{\mu}m\cos{\beta}\!=\!0\\
&s_{\mu}m\sin{\beta}\!+\!\nabla_{\mu}\ln{\phi}\!=\!0
\end{eqnarray}
the first solvable for the YT angle as
\begin{eqnarray}
&\nabla_{\mu}\ln{(\tan{\beta}\!+\!\sec{\beta})}\!=\!-2ms_{\mu}\label{YT}
\end{eqnarray}
which can be integrated and substituted into the second
\begin{eqnarray}
&\nabla_{\mu}\ln{\phi^{2}}\!=\!-2m\sin{\beta}s_{\mu}\label{module}
\end{eqnarray}
which then can also be eventually solved for the module.

The HJ and continuity equations are
\begin{eqnarray}
&\!\!\!\!\left|\nabla \frac{\beta}{2}\right|^{2}\!\!-\!m^{2}\!-\!\phi^{-1}\nabla^{2}\phi\!=\!0\\
&\!\nabla_{\mu}(\phi^{2}\nabla^{\mu}\frac{\beta}{2})\!=\!0
\end{eqnarray}
while the sine-KG and KG equations are
\begin{eqnarray}
&\nabla^{2}(2\beta)\!-\!4m^{2}\sin{(2\beta)}\!=\!0\\
&\nabla^{2}\phi^{2}\!+\!4m^{2}\phi^{2}\!=\!0
\end{eqnarray}
whose solutions can be found with the usual analysis.

Solutions (\ref{YT}, \ref{module}) can only be obtained if the tetradic structure of the space-time is assigned. However, in very formal ways, we can write a generic solution as
\begin{eqnarray}
&\beta\!=\!-\arcsin{[\tanh{(2m\!\int\!s_{\mu}dx^{\mu})}]}
\end{eqnarray}
so that
\begin{eqnarray}
&\phi\!=\!K\sqrt{\cosh{(2m\!\int\!s_{\mu}dx^{\mu})}}
\end{eqnarray}
which is known only when the spin content is also known.

Nonetheless, no spin is present in the HJ and continuity equations nor in the sine-KG and KG equations, but once solutions are found they have to be constrained by being plugged into the original Dirac field equations (\ref{YT}, \ref{module}).

We did not expect this solution to be physical because, after all, it was found in the free case, but we still can use it as the zero-order term of a full perturbative expansion in the coupling constants of some chosen interaction.

A more physical solution may be obtained in the interacting case, and we begin allowing torsion interactions.

The Dirac equations with pure torsion are
\begin{eqnarray}
&-XW_{\mu}\!+\!\nabla_{\mu}\beta/2\!+\!s_{\mu}m\cos{\beta}\!=\!0\\
&s_{\mu}m\sin{\beta}\!+\!\nabla_{\mu}\ln{\phi}\!=\!0
\end{eqnarray}
and from these it is straightforward to explicitly write
\begin{eqnarray}
&XW_{\mu}\!=\!\nabla_{\mu}\beta/2\!+\!s_{\mu}m\cos{\beta}
\label{torsion}
\end{eqnarray}
as the torsion axial-vector in terms of the YT angle.

As before, we proceed to write expressions
\begin{eqnarray}
&\!\!\!\!\left|\nabla \frac{\beta}{2}\right|^{2}\!\!-\!m^{2}\!-\!\phi^{-1}\nabla^{2}\phi
\!-\!X^{2}W^{2}\!=\!0\\
&\!\nabla_{\mu}(\phi^{2}\nabla^{\mu}\frac{\beta}{2})
\!-\!4m\frac{X^{2}}{M^{2}}\phi^{4}\sin{\beta}\!=\!0
\end{eqnarray}
as the HJ equation and the continuity equation, and
\begin{eqnarray}
&\nabla^{2}\beta\!-\!4m(m\cos{\beta}\!+\!XW^{\mu}s_{\mu}
\!+\!2\frac{X^{2}}{M^{2}}\phi^{2})\sin{\beta}\!=\!0\\
&\nabla^{2}\phi^{2}\!+\!4m(m\!+\!XW^{\mu}s_{\mu}\cos{\beta})\phi^{2}\!=\!0
\end{eqnarray}
as the sine--Klein-Gordon and Klein-Gordon equations.

When (\ref{torsion}) is injected into the torsion field equations
\begin{eqnarray}
&\nabla_{\alpha}(\partial W)^{\alpha\mu}\!+\!M^{2}W^{\mu}\!=\!2X\phi^{2}s^{\mu}
\end{eqnarray}
we obtain
\begin{eqnarray}
\nonumber
&(\nabla^{2}\cos{\beta}\!+\!M^{2}\cos{\beta})s^{\mu}m
\!-\!\nabla^{\mu}\nabla^{\alpha}\cos{\beta}s_{\alpha}m+\\
&+\frac{1}{2}M^{2}\nabla^{\mu}\beta
\!=\!2X^{2}\phi^{2}s^{\mu}\label{equation}
\end{eqnarray}
as a constraint among the components of the spinor field.

This field equation encompasses the field equations of torsion and one of the two Dirac field equations, whereas the other, being inverted according to
\begin{eqnarray}
&\beta\!=\!\arcsin{(\frac{1}{2m}s^{\mu}\nabla_{\mu}\ln{\phi^{2}})}
\end{eqnarray}
may then be substituted into the above (\ref{equation}), in order to get an equation in terms of the module: this equation as a matter of principle can be solved, but as a matter of fact it results into a differential equation that is not anything like an equation one would expect to solve exactly.

A different approach is to assume that the dynamics of the Dirac field be such that for a sufficiently large torsion mass, an effective approximation can be done, so that 
\begin{eqnarray}
&M^{2}W^{\mu}\!=\!2X\phi^{2}s^{\mu}
\end{eqnarray}
and, after integrating torsion, we end up with expression
\begin{eqnarray}
&2\frac{X^{2}}{M^{2}}\phi^{2}s_{\mu}\!=\!\nabla_{\mu}\beta/2\!+\!s_{\mu}m\cos{\beta}
\end{eqnarray}
as the first Dirac field equation; this can be plugged into the second Dirac field equation, giving one second-order differential field equation that, for small YT angle, reads
\begin{eqnarray}
&\nabla^{2}\phi\!-\!4m\frac{X^{2}}{M^{2}}\phi^{3}\!+\!2m^{2}\phi\!=\!0
\end{eqnarray}
as the equation of a soliton: solutions are given by
\begin{eqnarray}
&\phi\!=\!\sqrt{m}\frac{M}{X}[\cosh{(\sqrt{2}m\!\int\!s_{\mu}dx^{\mu})}]^{-1}
\end{eqnarray}
from which
\begin{eqnarray}
&\beta\!=\!\sqrt{2}\tanh{(\sqrt{2}m\!\int\!s_{\mu}dx^{\mu})}
\end{eqnarray}
which are respectively a soliton and a topological soliton, and of course they both solve the two original Dirac field equations within the limit of small YT angle and for the effective approximation of massive torsion field \cite{Fabbri:2016fxt}.

But even in regimes in which soliton equations can be obtained, no exact solution is known \cite{t,MacKenzie:2001av}.

In the more general case in which also electrodynamics is considered, the Dirac field equations (\ref{f1}, \ref{f2}) are
\begin{eqnarray}
&\!\!\!\!-(P\!-\!qA)^{\iota}u_{[\iota}s_{\mu]}\!-\!XW_{\mu}\!+\!\nabla_{\mu}\beta/2
\!+\!s_{\mu}m\cos{\beta}\!=\!0\\
&\!\!\!\!-(P\!-\!qA)^{\rho}u^{\nu}s^{\alpha}\varepsilon_{\mu\rho\nu\alpha}
\!+\!s_{\mu}m\sin{\beta}\!+\!\nabla_{\mu}\ln{\phi}\!=\!0
\end{eqnarray}
as it is easy to see; combining them one can work out
\begin{eqnarray}
\nonumber
&(P\!-\!qA)^{\nu}\!=\!m\cos{\beta}u^{\nu}
\!+\!s^{[\nu}u^{\mu]}(\frac{1}{2}\nabla_{\mu}\beta\!-\!XW_{\mu})+\\
&+\varepsilon^{\nu\rho\sigma\mu}s_{\rho}u_{\sigma}\nabla_{\mu}\ln{\phi}
\label{momentum}
\end{eqnarray}
as the expression of the momentum of the Dirac field.

It is insightful to compare this form to that of \cite{r-w}.

The Hamilton-Jacobi equation (\ref{HJ}) reduces to
\begin{eqnarray}
\nonumber
&\!\!\frac{q}{2}F^{\mu\nu}s^{\alpha}u^{\sigma}\varepsilon_{\mu\nu\alpha\sigma}\!+\!(P\!-\!qA)^{2}
\!-\!X(P\!-\!qA)^{[\iota}W^{\mu]}u_{[\iota}s_{\mu]}-\\
&-X^{2}W^{2}\!+\!|\nabla \beta/2|^{2}\!-\!\phi^{-1}\nabla^{2}\phi\!-\!m^{2}\!=\!0
\end{eqnarray}
where it becomes clear the role of the YT angle in being a corrective term to the quantum potential \cite{r-w,m-r-w}.

A connection of this expression to the de Broglie-Bohm theory was also discussed in \cite{h-c}. A controversy between the results of \cite{r-w,m-r-w} and \cite{h-cA, h-cB} has also been portrayed but we are not going to deepen it in the present work.

From (\ref{momentum}) one may further take the curl obtaining
\begin{eqnarray}
\nonumber
&\!\!qF^{\alpha\nu}\!=\!m\sin{\beta}u^{[\nu}\nabla^{\alpha]}\beta
\!+\!s^{\mu}u^{[\nu}\nabla^{\alpha]}(\frac{1}{2}\nabla_{\mu}\beta\!-\!XW_{\mu})-\\
&\!\!\!\!-u^{\mu}s^{[\nu}\nabla^{\alpha]}(\frac{1}{2}\nabla_{\mu}\beta\!-\!XW_{\mu})
\!+\!s_{\rho}u_{\sigma}\varepsilon^{\mu\rho\sigma[\nu}\nabla^{\alpha]}\nabla_{\mu}\ln{\phi}
\end{eqnarray}
as the electrodynamic strength of the Dirac field that is to be substituted into the electrodynamic equations
\begin{eqnarray}
&\nabla_{\sigma}F^{\sigma\mu}\!=\!2q\phi^{2}u^{\mu}
\end{eqnarray}
so that one obtains
\begin{eqnarray}
\nonumber
&m\cos{\beta}u^{[\nu}\nabla^{\alpha]}\beta\nabla_{\alpha}\beta
\!+\!m\sin{\beta}u^{[\nu}\nabla^{\alpha]}\nabla_{\alpha}\beta+\\
\nonumber
&+s^{\mu}u^{[\nu}\nabla^{\alpha]}\nabla_{\alpha}(\frac{1}{2}\nabla_{\mu}\beta\!-\!XW_{\mu})-\\
\nonumber
&-u^{\mu}s^{[\nu}\nabla^{\alpha]}\nabla_{\alpha}(\frac{1}{2}\nabla_{\mu}\beta\!-\!XW_{\mu})+\\
&+s_{\rho}u_{\sigma}\varepsilon^{\mu\rho\sigma[\nu}\nabla^{\alpha]}
\nabla_{\alpha}\nabla_{\mu}\ln{\phi}
\!=\!2q^{2}\phi^{2}u^{\nu}
\end{eqnarray}
as a constraint among the degrees of freedom of the Dirac field, given by module and YT angle coupled to torsion.

After this equation is solved, we get the structural form of the electrodynamic strength produced by an assigned distribution of torsionally-interacting spinor fields; when torsion is negligible, the electrodynamic force is produced by the spinor distribution alone, and our results become comparable to those of \cite{c-c-r-b}: in this reference, the authors study the Dirac equation in presence of electrodynamics, discussing what they call ``Relativistic Dynamical Inversion'', a method in terms of which, starting from the Dirac equation in presence of electrodynamics, we may invert it to get the electrodynamic potential in terms of the spinor field itself. Their method is general, but eventually they need perform a consistency check on the electrodynamic potential to see if it is Hermitian; as (\ref{momentum}) clearly shows, the electrodynamic potential is in fact always Hermitian, and henceforth real and physical. These types of method can be very interesting, not only when looking for exact solutions, but also in a lot of applications stretching from lasers, optics, trapped ions, cold atoms, circuit quantum electrodynamics, relativistic quantum chemistry, to some general quantum technologies like information processing as it has been discussed in \cite{c-c-r-b} and references therein.

It has been discussed how RDI can also be generalized to other interactions, including scalar potentials coupled to the mass, and this is also true for the method we have presented here above, in which we have included torsional interactions, and where gravity may be added as well.

In fact, when gravitation is present, problems like the stability and localization of the module assume entirely different forms: for example, in pure gravitational cases and setting the YT angle to be equal to zero, we have
\begin{eqnarray}
&\frac{1}{2}\varepsilon_{\mu\alpha\nu\iota}(R\!-\!\Omega)^{\alpha\nu\iota}\!+\!2s_{\mu}m\!=\!0\\
&(R\!-\!\Omega)_{\mu a}^{\phantom{\mu a}a}\!+\!\nabla_{\mu}\ln{\phi^{2}}\!=\!0
\end{eqnarray}
showing that the dual of the gravitational field, namely the curl of the tetrads, is related to the mass, through the spin axial-vector, while the trace of the gravitational field, namely the divergence of the tetrads, determines a profile of the module, which can be localized and stable.

However, the module is not the main concern we have now because it is much more interesting to assess the role that is played by the ubiquitous but elusive YT angle.

In what remains to be done, we will give a quantitative discussion about this mysterious pseudo-scalar field.
%%%%%%%%%%%%%%%%%%%%%%%%%%%%%%%%%%%%%%%%%%%%%%%%%%%%%%%%%%%%%%%%%%%%%%%%%%%%%%%%%%%%%%%%%%%%%%%%%%%
%%%%%%%%%%%%%%%%%%%%%%%%%%%%%%%%%%%%%%%%%%%%%%%%%%%%%%%%%%%%%%%%%%%%%%%%%%%%%%%%%%%%%%%%%%%%%%%%%%%
\section{General Considerations}
We have seen that the spinor field in general possesses a total number of $8$ real components, given by the $3$ components of the velocity and the $3$ components of the spin, a module and the YT angle (\ref{b2}, \ref{b1}): these last two being the only true degrees of freedom. We have also seen that the Dirac spinor field equations are $8$ real equations that are equivalent to the two vector equations (\ref{f1}, \ref{f2}) giving all the derivatives of these two scalar fields. From these, we have studied some specific situations in mathematical terms, and although the results we found were either too difficult to treat or too simple to be realistic, nevertheless there is some information we can extract in order to give a meaning to various fields, especially to the YT angle.

In the form (\ref{spinor}), taken for small velocities, we have
\begin{eqnarray}
&\!\psi\!=\!\phi e^{i\alpha}\left(\!\begin{tabular}{c}
$e^{\frac{i}{2}\beta}\left(\boldsymbol{\mathbb{I}}\!-\!
\vec{v}\!\cdot\!\vec{\frac{\boldsymbol{\sigma}}{2}}\right)\xi$\\
$e^{-\frac{i}{2}\beta}\left(\boldsymbol{\mathbb{I}}\!+\!
\vec{v}\!\cdot\!\vec{\frac{\boldsymbol{\sigma}}{2}}\right)\xi$
\end{tabular}\!\right)
\label{ch}
\end{eqnarray}
showing that left-handed and right-handed semi-spinorial fields are distinguished in terms of their YT angle, and so the YT angle is what keeps the two chiral projections independent even in the rest frame. Like we have discussed in reference \cite{Fabbri:2016msm}, the non-relativistic limit requires small spatial part of the velocity vector as well as a small YT angle, indicating that if the velocity vector describes the overall motion then the YT angle describes some sort of internal motion, which we may define as that motion that remains even in the rest frame; in complementary cases of ultra-relativistic limit, the YT angle also vanishes: thus, the YT angle vanishes whenever the two chiral parts are either equal (as in the non-relativistic case) or totally separable (as in the ultra-relativistic case), confirming that the YT angle encodes information about relative motions of the two chiral parts. In the standard representation, for small velocities as well as small YT angle, we have
\begin{eqnarray}
&\!\psi\!=\!\phi e^{i\alpha}\sqrt{2}\left(\!\begin{tabular}{c}
$\xi$\\
$-\frac{1}{2}\left(i\beta\!-\!\vec{v}\!\cdot\!\vec{\boldsymbol{\sigma}}\right)\xi$
\end{tabular}\!\right)
\label{st}
\end{eqnarray}
showing in what way the YT angle is linked to the small and large semi-spinors: then the large semi-spinor may be interpreted as the average of the two chiral parts whereas the small semi-spinor may be interpreted as the standard deviation of the two chiral parts, and thus again the YT angle appears to dictate the way in which the two chiral parts deviate from mean-field configurations. YT angles that are trivial would be those of a spinor field possessing no degree of freedom intrinsic to the matter distribution.

This is the case for instance in QFT treated with perturbative methods: in this case in fact, plane waves are employed, and plane-wave solutions do have a vanishing YT angle. On the other hand, another quantity that for plane-wave states is zero is the so-called \textit{zitterbewegung}, or jittering motion, the trembling motion of the spinorial particles. Could there be, therefore, any relationship?

Another qualitative hint may come from the fact that, as discussed in \cite{Fabbri:2016laz}, the presence of the YT angle and the dynamics it induces on spin may be responsible for effects similar to those usually attributed to field quantization, and the same parallel has also been discussed in \cite{h4} as well as \cite{Recami:1995iy,Salesi:1995vy} in connection to the \textit{zitterbewegung} effect.

So could there really be some relation between the YT angle and \textit{zitterbewegung} effects on the particle?

To render the argument more quantitative, we might notice that \textit{zitterbewegung} in its commonly accepted form can only happen if momentum and velocity are not proportional: more in detail, let us consider (\ref{momentum}) in the case of no interactions, for which we get the simple form
\begin{eqnarray}
\nonumber
&P^{\nu}\!=\!m\cos{\beta}u^{\nu}\!+\!\frac{1}{2}\nabla_{\mu}\beta u^{[\mu}s^{\nu]}+\\
&+\varepsilon^{\nu\rho\sigma\mu}s_{\rho}u_{\sigma}\nabla_{\mu}\ln{\phi}
\end{eqnarray}
showing that the YT angle acts in terms of its cosine to change the length, and where both YT angle and module appear in a derivative form, as source of spin divergence, to change the direction: $\sin{\beta}\!=\!0$ and a constant module together imply that $P^{\nu}\!=\!mu^{\nu}$ so no \textit{zitterbewegung} could occur; then no \textit{zitterbewegung} means $P^{0}\vec{u}\!=\!\vec{P}$ and in turn this means the constancy of module and YT angle, as it can be seen by checking order by order in the spatial part of the velocity following a perturbative analysis. Hence, what is essential for the \textit{zitterbewegung} is the dynamics of both the YT angle and the module, and as a consequence, the YT angle is only partially related to the mechanism that brings about the \textit{zitterbewegung} because even with no YT angle there is still \textit{zitterbewegung} so long as there is a module displaying its own proper dynamical character.

Nonetheless, in situations where the module is constant the dynamics of the YT angle becomes necessary in order to see the appearance of \textit{zitterbewegung} effects, as should have been clear from the fact that in these situations the appearance of effects of \textit{zitterbewegung} is linked to the appearance of the small semi-spinor: as (\ref{st}) shows in a very clear manner, small semi-spinorial components may be present if the YT angle is present. And this situation also occurs even if we are in the frame that is at rest.

Since from Dirac field equations (\ref{f1}) the dynamics of the YT angle is linked to the presence of torsion, then we may wonder what are the effects of torsion for the general mechanics that involves \textit{zitterbewegung} phenomena.

We will leave this problem to following works.
%%%%%%%%%%%%%%%%%%%%%%%%%%%%%%%%%%%%%%%%%%%%%%%%%%%%%%%%%%%%%%%%%%%%%%%%%%%%%%%%%%%%%%%%%%%%%%%%%%%
%%%%%%%%%%%%%%%%%%%%%%%%%%%%%%%%%%%%%%%%%%%%%%%%%%%%%%%%%%%%%%%%%%%%%%%%%%%%%%%%%%%%%%%%%%%%%%%%%%%
\section{Conclusion}
In this paper, we have considered the polar decomposition of the spinor field applied to the spinor field equations, that is the Dirac equations, decomposing them into various equations that are real and written in terms of tensors alone, known to be the Madelung-Gordon decompositions; when this was done, we proceeded in selecting two of these, showing that they imply all remaining ones, and in fact proving that these two are equivalent to the initial polar form of spinor field equations, and namely to the Dirac equations: therefore, we can say that the Dirac field equations are equivalent to (\ref{f1}, \ref{f2}). We discussed how the polar form of the spinor shows that, among all spinor components, only two are degrees of freedom, and namely, the module and the YT angle: as a consequence, the fact that the Dirac spinor field equations are equivalent to a pair of vectorial equations is obvious, since two vectorial equations are exactly what determines all of the derivatives of two real scalar fields. We find this circumstance to be very interesting, as it is important to possess some formal simplification for the Dirac equations.

On the other hand, we have seen that, despite the formal simplification brought by having the Dirac equations written equivalently as (\ref{f1}, \ref{f2}), nonetheless the last two equations are still far from being easy to manipulate, and we have only managed to find some avenue leading to exact solutions, but without having been able to actually follow them till the end; this is unfortunate. However, we found some application to technological methodologies in which these equations did bring relevant advantages.

Finally, we discussed how these two equations, taking into account the analysis of \cite{Fabbri:2016msm} on non-relativistic limits, could be related to the \textit{zitterbewegung} phenomena which, in light of results such as those in \cite{Welton:1948zz} on the Lamb shift, might well make the use of the present methods rather intriguing also in domains of pure theoretical interest.

Whether it is for fundamental results or practical applications, it should be quite clear what is the advantage of having a simpler form of the Dirac equations.

Here we presented the simplest of them all.
%%%%%%%%%%%%%%%%%%%%%%%%%%%%%%%%%%%%%%%%%%%%%%%%%%%%%%%%%%%%%%%%%%%%%%%%%%%%%%%%%%%%%%%%%%%%%%%%%%%
\begin{acknowledgments}
I wish to warmly thank Professor Rold\~{a}o da Rocha for the discussions from which I have had the essential idea that is at the core of the present article.
\end{acknowledgments}
%%%%%%%%%%%%%%%%%%%%%%%%%%%%%%%%%%%%%%%%%%%%%%%%%%%%%%%%%%%%%%%%%%%%%%%%%%%%%%%%%%%%%%%%%%%%%%%%%%%
%%%%%%%%%%%%%%%%%%%%%%%%%%%%%%%%%%%%%%%%%%%%%%%%%%%%%%%%%%%%%%%%%%%%%%%%%%%%%%%%%%%%%%%%%%%%%%%%%%%

%%%%%%%%%%%%%%%%%%%%%%%%%%%%%%%%%%%%%%%%%%%%%%%%%%%%%%%%%%%%%%%%%%%%%%%%%%%%%%%%%%%%%%%%%%%%%%%%%%%
\end{document}